\documentclass[11pt,a4paper]{article}
\usepackage{amsmath,amsfonts,amssymb,amscd}
\usepackage{graphicx}
\usepackage{epsfig}

\def\S{\mathcal S\/}

\def\vth{\vartheta}
\def\vph{\varphi}
\def\={\, = \,}

\begin{document}

\title{Conditions for
the feasibility of multiple rolling for mechanical systems with
multiple contact points}

\author{Stefano Pasquero \\
        Dipartimento di Matematica dell'Universit\`a di Parma \\
        Via G.P. Usberti, 53/A - 43100 Parma (Italia) \\
        E-mail: stefano.pasquero@unipr.it
        }
\date{}
\maketitle

\begin{abstract}
\noindent We illustrate a theoretical procedure determining
necessary conditions for which simultaneous pure rolling kinetic
constraints acting on a mechanical system can be fulfilled. We
also analyze the sufficiency of these conditions by generalizing
to this case a well known and usually accepted assumption on the
behavior of pure rolling constraint. We present in detail the
application of the procedure to some significative mechanical
systems.




\end{abstract}


\section*{Introduction}
Mechanical systems subject to rolling kinetic constraints are one
of the most studied argument of Classical Mechanics, especially
for its wideness of applicability in several branches of
Mechanical Sciences: Contact Mechanics, Tribology, Wear, Robotics,
Ball Bearing Theory and Control Theory applied to moving engines
and vehicles are only some of the important fields where the
results about pure rolling constraint can be fruitfully used.

It is well known that, when a mechanical system moves in contact
with an assigned rough surface, the effective fulfilment of the
kinetic conditions determined by the rolling without sliding
requirement of the system on the surface depends on the behavior,
with respect to the considered law of friction, of the reaction
forces acting on the system in the contact points. For example,
the roll of a disk on a rough straight line, considering the
Coulomb's law of friction, can happen only if the contact force
lie inside the friction cone (see Example 1 below).

However, even in the simplest case of a mechanical system formed
by a single rigid body, in the case of multiple contact points
between the rigid body and the rough surface, it could be an hard
task to obtain sufficient information about the contact reactions
in order to determine if the laws of friction are satisfied or not
during the motion. In fact the most common methods to determine
information about the reactions, starting from the simple
application of linear and angular momenta equations (see e.g.
\cite{LeviCivita,Goldstein}) to most refined techniques such as
lagrangian multipliers in lagrangian mechanics (see e.g.
\cite{Huston1999}) or deep analyses of the contact between the
system and the surface (see e.g. \cite{DeMoerlooze2011}), have a
global character. Then these methods, for their very nature, can
determine only a reactive force system equivalent to the real one
but, in the general case, these methods cannot determine the
single reactive forces in the contact points. The problem becomes
even more complicated in case of multibody system, due to the
presence of the internal reactions in the link between the parts
of the system.

In this paper we consider the motion of a mechanical system having
two or more distinct contact points with one or more assigned
rough surfaces, and we determine necessary conditions for which in
all the contact points the pure rolling kinetic constraint can
hold. We also analyze the sufficiency of these conditions by
generalizing to this case a well known and usually accepted
assumption on the behavior of pure rolling constraint. Moreover,
we briefly discuss the possible behaviors of the system when the
necessary conditions are not fulfilled.

The procedure to determine if the rolling condition can be
fulfilled can be applied both to systems formed by a single rigid
body and to multibody systems. It is essentially based on the
application of linear and angular momenta equations to the (parts
forming the) mechanical system, and therefore it gives an
underdetermined system in the unknown single contact reactions.
Nevertheless, we show that the lack of complete knowledge of the
single contact reactions is not an obstacle to determine the
feasibility of the rolling conditions.

It is however important to remark that, although the procedure has
a very simple and unassailable theoretic foundation, its effective
application to general systems could present insurmountable
difficulties. This is essentially due to the fact that the general
procedure explicitly requires the knowledge of the motion law of
the system, and in the general case the explicit time--dependent
expression of the motion cannot be obtained because of
complications determined by the geometry of the system itself
and/or by the integrability of the equations of motion.
Nevertheless there are several significative cases where the
procedure can be explicitly performed. In the paper, we illustrate
three examples with rising complication: the well known case of a
disk falling in contact with an inclined plane (that is presented
only to point out some key points of the general procedure); the
case of a system formed by a non--coupled pair of disks connected
with a bar and moving on the horizontal plane; the case of a heavy
sphere falling in contact with a guide having the form of a
V--groove non symmetric with respect to the vertical axis and
inclined with respect to the horizontal.

The main content of this paper can be approached starting from a
very standard background knowledge, essentially focused to the
linear and angular equations of motion for a mechanical system,
the so called Cardinal Equations, and the basic theory of pure
rolling conditions and kinetic constraints. On the other hand, the
list of possible references involving theory and application of
pure rolling constraint is almost endless. Therefore we chose to
cite only a very limited list of references sufficient to make the
paper self--consistent: the classical book of Levi--Civita and
Amaldi \cite{LeviCivita} and the book of Goldstein
\cite{Goldstein} for the Cardinal Equations and the basic concepts
about pure rolling conditions; the book of Neimark and Fufaev
\cite{NeimarkF} and the paper of Massa and Pagani
\cite{MassPaga1991} for the behavior of systems subject to kinetic
constraints. The interested reader can find in the wide but not
exhaustive lists of references of \cite{Brogliato,Johnson} as a
useful starting point to delve in the expanse of the material
related to this argument.

The paper is divided in four sections. Section 1 contains a very
brief preliminary description of the well known analysis of the
rolling condition for a disk in contact with an inclined plane.
This remind is motivated by some useful affinities with the
general procedure for generic systems. Section 2 contains the
discussion of the general case, and the determination of the
necessary conditions for pure rolling conditions simultaneously
hold. Section 3 presents the example of the system formed by the
non--coupled disks and the example of the heavy sphere falling in
the V--groove. Section 4 is devoted to open problems, remarks and
conclusions.

\section{Preliminaries}

\subsection*{Example 1.}
An homogeneous disk of mass $m$ and radius $R$ moves in the
vertical plane being in contact with a rough guide inclined with
slope angle $\alpha$.
\begin{figure}[h] \label{disco}
\centering \includegraphics[width=0.65\textwidth]{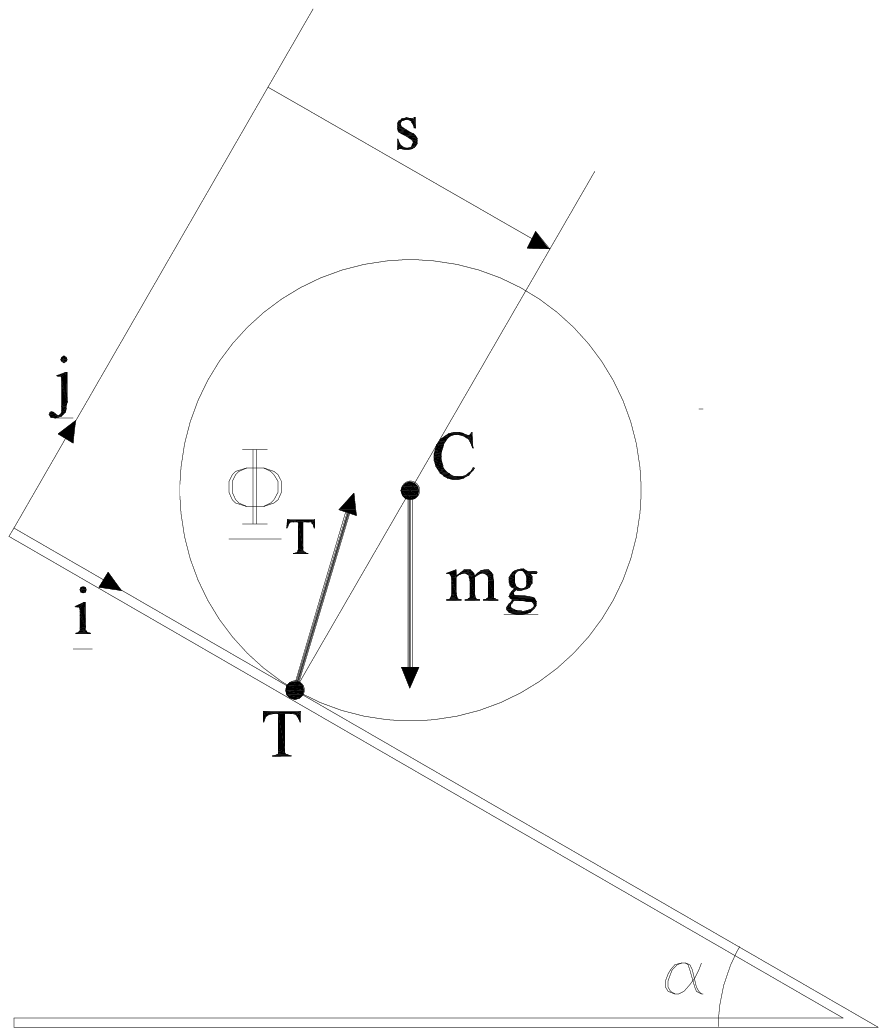}
\caption{Rolling disk on an inclined plane}
\end{figure}
Considering the system subject to the Coulomb's law of friction,
with obvious notation clarified by Fig. \ref{disco}, the
feasibility of pure rolling condition of the disk can be
determined with the following procedure:

\begin{itemize}

\item[0)] we determine the relative velocity $\underline{{\bf
v}}_T(t_0)$ of the contact point $T$ of the disk at the instant
$t_0$ with respect to the inclined plane as function of the
initial data of the motion. The pure rolling condition requires of
course that $\underline{{\bf v}}_T(t_0)=0$. If so

\item[1)] we assume that the disk rolls without sliding on the
inclined plane. Then the system has a single degree of freedom
(for example the coordinate $s$ of $T$ along the inclined plane)
and we can determine the equation of motion
\begin{equation*}
mg \sin \alpha \= \dfrac32 m \ddot{s} \, ;
\end{equation*}

\item[2)] we determine the corresponding reaction $\underline{{\bf
\Phi}}_T$ as a (in this case constant) function of time
\begin{equation*}
\underline{{\bf \Phi}}_T \= m \underline{{\bf a}}_C - m
\underline{{\bf g}} \= \left( - \dfrac13 m g \sin \alpha \right)
\underline{{\bf i}} + \left( mg \cos \alpha \right)
\underline{{\bf j}} \, ;
\end{equation*}

\item[3)] we test the Coulomb's law of friction condition
\begin{equation*}\label{condizioni_CM_es0}
\| \underline{{\bf \Phi}}_T^{\|} \| \le \mu \, \| \underline{{\bf
\Phi}}_T^{\perp} \| \quad \Leftrightarrow \quad \mu \, \ge
\dfrac13 \tan \alpha
\end{equation*}
where $\underline{{\bf \Phi}}_T^{\|}, \underline{{\bf
\Phi}}_T^{\perp}$ are the parallel and orthogonal component of
$\underline{{\bf \Phi}}_T$ with respect to the inclined plane;

\item[4)] we assume that, if and until the Coulomb's condition is
verified, the disk moves rolling on the plane and that  if and
when the Coulomb's condition is not verified, the disk changes its
dynamic evolution beginning to slide on the plane (until the first
time $t_1
> t_0$ such that $\underline{{\bf v}}_T(t_1)=0$).

\end{itemize}

Some remarks are in order to focus the possibility to generalize
the procedure to more complicated systems. Step 2 consists in the
determination of the reaction acting on the disk as function of
time. The utmost simplicity of the specific problem can hide the
fact that in a more general situation the information about the
reaction sufficient to analyze the rolling condition could require
an explicit determination of the motion of the system as function
of time.

Step 3 tests the compatibility of the reaction evaluated in Step 2
with the Coulomb's law of friction assumed as the constitutive
characterization of the rough surface in contact with the disk. Of
course the feasibility of the rolling condition can be tested with
any other significative constitutive law.

In Step 4 we assume that, roughly speaking, if the disk can roll
then it does. This is of course an arbitrary assumption, but the
hypothesis is well confirmed by experimental results. In the next
section, we will confirm this assumption in the more general
situation of generic system.

To conclude the section, let us note that, in this very simple
case, both the behaviors of the disk when the Coulomb's friction
condition is or is not verified are determinable. In the general
case, when the constitutive law is not verified, the behavior of
the system turns out to be not so straight to determine, although
some reasonable assumptions can be done. We will go back on this
arguments in Section 4.

\section{The general case}

In this section, following a line of though similar to the one
applied in the previous section, we discuss the possibility that a
mechanical system $\S$ having two points $T_1,T_2$ in contact with
a fixed surface $\Sigma$ moves such that in both the contact
points the rolling conditions can subsist respecting the Coulomb's
law of friction. The arguments of the discussion can be easily
extended to cases with more (but a finite number) than two contact
points and possibly to different friction constitutive laws.

The discussion is based on the fact that, along the motion, the
reactive forces acting on the system must validate the linear and
angular momenta equations
\begin{eqnarray}\label{sistema_generale}
\left\{
\begin{array}{l}
\underline{{\bf R}}^{act} + \underline{{\bf R}}^{react} \= M
\underline{{\bf a}}_G \\ \\

\underline{{\bf M}}_G^{act} + \underline{{\bf M}}_G^{react} \=
\dfrac{d \, \underline{\Gamma}_G}{d t}
\end{array}
\right.
\end{eqnarray}
where $M$ is the total mass of the system, $G$ is the center of
mass of the system and $\underline{{\bf R}}^{act}, \underline{{\bf
R}}^{react},\underline{{\bf M}}_G^{act},\underline{{\bf
M}}_G^{react}$ are respectively the sum of the active and reactive
forces and active and reactive momenta acting on the whole system.
In this specific situation we have that:
\begin{eqnarray}\label{determinazione_reazioni}
\left\{
\begin{array}{l}
\underline{{\bf R}}^{react} \= \underline{{\bf \Phi}}_{T_1} +
\underline{{\bf \Phi}}_{T_2}
\\ \\
\underline{{\bf M}}_G^{react} \= \overrightarrow{G {T_1}} \times
\underline{{\bf \Phi}}_{T_1} + \overrightarrow{G {T_1}} \times
\underline{{\bf \Phi}}_{T_2}
\end{array}
\right. .
\end{eqnarray}
It is however well known \cite{LeviCivita} that Eqs.
(\ref{sistema_generale}) are not sufficient to determine the
motion of the mechanical system and the single reactions
$\underline{{\bf \Phi}}_{T_1}, \underline{{\bf \Phi}}_{T_2}$ along
the motion, since the system
\begin{eqnarray}\label{sistema_sottodeterminato}
\left\{
\begin{array}{l}
\underline{{\bf R}}^{act} + \underline{{\bf \Phi}}_{T_1} +
\underline{{\bf \Phi}}_{T_2} \= M
\underline{{\bf a}}_G \\ \\

\underline{{\bf M}}_G^{act} + \overrightarrow{G{T_1}} \times
\underline{{\bf \Phi}}_{T_1} + \overrightarrow{G{T_2}} \times
\underline{{\bf \Phi}}_{T_2} \= \dfrac{d \,
\underline{\Gamma}_G}{d t}
\end{array}
\right.
\end{eqnarray}
is by its very nature under--determined. In fact the projection of
the angular momenta equation of (\ref{sistema_sottodeterminato})
in the direction of $\overrightarrow{{T_1}{T_2}}$ is a pure
equation of motion of the system where no reactions appear. Then
(\ref{sistema_sottodeterminato}) can give no more than 5 relations
on the components of $\underline{{\bf \Phi}}_{T_1}$ and
$\underline{{\bf \Phi}}_{T_2}$. Unfortunately, due to the
roughness of the contacts, no preliminary conditions can be
imposed on the components of the reactions, so that, even when the
motion of the mechanical system is known,
(\ref{sistema_sottodeterminato}) is a linear system with $6$
unknowns that is not of maximum rank.

Nevertheless the parametric solution of the system
(\ref{determinazione_reazioni}), and an assumption parallelizing
the one of Step 4 of the case of Section 1, give us the
possibility of determining if the rolling conditions in $T_1$ and
$T_2$ are or not verified.

The procedure to test the feasibility of pure rolling condition of
the disk can be then based on the following steps:
\begin{itemize}

\item[0)] we test if the initial relative velocities of the
contact points $T_1,T_2$ with respect to the surface are null or
not. If they are null

\item[1)] we suppose that the system rolls without sliding in both
the contact points. This assumption fixes the dynamics (for
example the number of degrees of freedom...) of the system and
consequently allows the determination of the motion of the system;

\item[2)] we write the linear and angular momenta equations for
the whole system, for example in the form:
\begin{eqnarray}\label{sistema_reazioni}
\left\{
\begin{array}{l}
\underline{{\bf \Phi}}_{T_1} + \underline{{\bf \Phi}}_{T_2} \= M
\underline{{\bf a}}_G - \underline{{\bf R}}^{act} \\ \\

\overrightarrow{G{T_1}} \times \underline{{\bf \Phi}}_{T_1} +
\overrightarrow{G{T_2}} \times \underline{{\bf \Phi}}_{T_2} \=
\dfrac{d \, \underline{\Gamma}_G}{d t} - \underline{{\bf
M}}_G^{act}
\end{array}
\right. .
\end{eqnarray}
Since the motion of the system is known, both the right hand sides
of the equations, together with the position vectors
$\overrightarrow{G{T_1}}$ and $\overrightarrow{G{T_2}}$, are known
as function of time. Therefore Eqs. (\ref{sistema_reazioni}) turn
out to be a time--dependent under--determined linear system in the
six scalar unknowns given by the components of the vectors
$\underline{{\bf \Phi}}_{T_1},\underline{{\bf \Phi}}_{T_2}$;

\item[3)] we solve the linear under--determined system
(\ref{sistema_reazioni}), obtaining the expression of the reaction
$\underline{{\bf \Phi}}_{T_1}$ and $\underline{{\bf \Phi}}_{T_2}$
as function of time and parameters $\lambda_1,\dots,\lambda_r$,
where of course the integer $r$ is related to the rank of
(\ref{sistema_reazioni}). Then we can determine the tangent and
orthogonal components $\underline{{\bf \Phi}}_{T_1}^{\|},
\underline{{\bf \Phi}}_{T_1}^{\perp}, \underline{{\bf
\Phi}}_{T_2}^{\|}, \underline{{\bf \Phi}}_{T_2}^{\perp}$ of the
reactions with respect to the surface $\Sigma$ as functions of
$(t,\lambda_1,\dots,\lambda_r)$. The pure rolling conditions then
can subsist in both the contact points only in the time interval
$[t_0, t_1]$ such that for every $t\in [t_0, t_1]$ there exists at
least one admissible $r$--uple
$(\overline{\lambda}_1,\dots,\overline{\lambda}_r)$ such that the
system
\begin{eqnarray}\label{condizioni_CM}
\left\{
\begin{array}{l}
\| \underline{{\bf \Phi}}_{T_1}^{\|}(t,
\overline{\lambda}_1,\dots,\overline{\lambda}_r) \| \le \mu_1 \,
\| \underline{{\bf \Phi}}_{T_1}^{\perp}(t,
\overline{\lambda}_1,\dots,\overline{\lambda}_r)\|
\\ \\
\| \underline{{\bf
\Phi}}_{T_2}^{\|}(t,\overline{\lambda}_1,\dots,\overline{\lambda}_r)
\| \le \mu_2 \, \|\underline{{\bf
\Phi}}_{T_2}^{\perp}(t,\overline{\lambda}_1,\dots,\overline{\lambda}_r)\|
\end{array}
\right.
\end{eqnarray}
holds;

\item[4)] we assume that, if for every $t\in [t_0, t_1]$ there
exists at least one admissible $r$--uple
$(\overline{\lambda}_1,\dots,\overline{\lambda}_r)$ such that
(\ref{condizioni_CM}) are verified, the system moves rolling
without sliding in both points $T_1$ and $T_2$ during the time
interval $[t_0, t_1]$.

\end{itemize}

It is clear that the general procedure described above
parallelizes as possible and generalizes the one of the disk on
the inclined plane. The most significant differences consist in
the  explicit determination of the motion of the system (since
otherwise Eqs. (\ref{sistema_reazioni}) could not admit a simple
parametric solution for the reactions $\underline{{\bf
\Phi}}_{T_1}$ and $\underline{{\bf \Phi}}_{T_2}$) and in the fact
that, when the Coulomb conditions (\ref{condizioni_CM}) are NOT
verified, being understood that the system does not roll in both
contact points, the determination of the behavior of the system
could require a more subtle analysis. We will go back on these
arguments in Section 4. We also remark that not all the $r$--uple
$\lambda_1,\dots,\lambda_r$ could be admissible in the discussion
of the inequalities (\ref{condizioni_CM}). For example, if the
system is leaned on the surface, we have to restrict our attention
to the $r$--uple such that
\begin{eqnarray}\label{condizioni_appoggio}
\left\{
\begin{array}{l}
\underline{{\bf \Phi}}_{T_1}^{\perp}(t, \lambda_1,\dots,\lambda_r)
\cdot \underline{{\nu}}_1 \ge 0
\\ \\
{\underline{{\bf
\Phi}}_{T_2}^{\perp}(t,\lambda_1,\dots,\lambda_r)} \cdot
\underline{{\nu}}_2 \ge 0
\end{array}
\right.
\end{eqnarray}
(where $\underline{{\nu}}_i$ is the unit normal vector to the
surface $\Sigma$ in the point $T_i$ and orientated toward the side
of the system) since otherwise the system detaches from the
surface.

\section{Examples}

\begin{figure}[h] \label{ruote}
\centering
\includegraphics[width=0.85\textwidth]{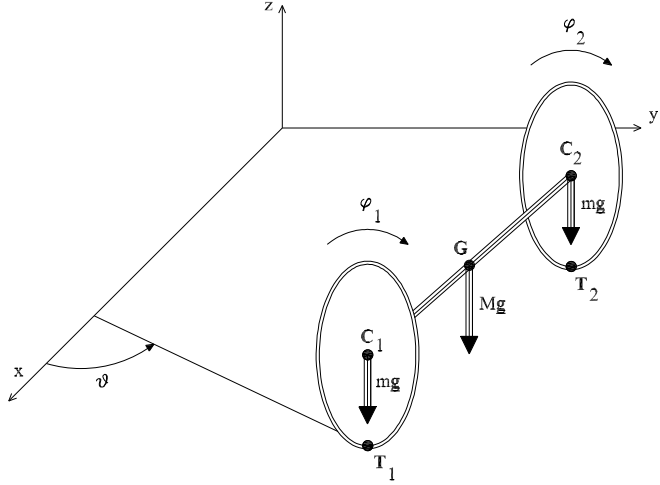}
\caption{Rolling system on an horizontal plane}
\end{figure}

A mechanical system is formed by two equal disks of mass $m$ and
radius $R$ and a rod, of mass $M$ and length $L$. The rod is
constrained to remain orthogonal to the two planes of the disks
with its endpoints coinciding with the two centers of the disks
(see Fig. \ref{ruote}) so that the disks remain vertical. The
whole system is leaned on a rough horizontal plane. The system has
then $5$ degrees of freedom: the coordinates $x,y$ of the center
of mass $G$ of the rod, the angle $\vth$ formed by the plane of
the disks with the $xz$ plane and the two rotation angles $\vph_1,
\vph_2$ of the disks. The rolling conditions in the contact points
$T_1$ and $T_2$ are equivalently expressed by:
\begin{eqnarray}\label{condizioni_PRes1}
\hskip -1truecm \left\{
\begin{array}{l}
\dot{x} + \dfrac12 L \dot{\vth} \cos \vth - R \dot{\vph_1} \cos
\vph_1 \=0
\\ \\
\dot{y} + \dfrac12 L \dot{\vth} \sin \vth - R \dot{\vph_1} \sin
\vph_1 \=0
\\ \\
\dot{\vth} - \dfrac{R}{L} \left(\dot{\vph_1} - \dot{\vph_2}
\right) \=0
\end{array}
\right. \quad \Leftrightarrow \quad \left\{
\begin{array}{l}
\dot{x} - \dfrac12 L \dot{\vth} \cos \vth - R \dot{\vph_2} \cos
\vph_2 \=0
\\ \\
\dot{y} - \dfrac12 L \dot{\vth} \sin \vth - R \dot{\vph_2} \sin
\vph_2 \=0
\\ \\
\dot{\vth} - \dfrac{R}{L} \left(\dot{\vph_1} - \dot{\vph_2}
\right) \=0
\end{array}
\right.
\end{eqnarray}
Tedious but straightforward computations (see
\cite{NeimarkF,MassPaga1991}) give the equations of motion of the
system
\begin{eqnarray}\label{eq_moto_es1}
\left\{
\begin{array}{lcl}
\ddot{x} &=& \dfrac{1}{2} L \dot{\vth}^2 \sin \vth - R
\dot{\vth}\dot{\vph_1} \sin \vth
\\ \\
\ddot{y} &=& - \dfrac{1}{2} L \dot{\vth}^2 \cos \vth + R
\dot{\vth}\dot{\vph_1} \cos \vth
\\
\ddot{\vth} &=& 0
\\
\ddot{\vph_1} &=& 0
\\
\ddot{\vph_2} &=& 0
\end{array}
\right.
\end{eqnarray}
If we suppose assigned the almost generic initial data
\begin{eqnarray}\label{dati_iniziali_es1}
\left\{
\begin{array}{lcl}
x(0) &=& x_0
\\
y(0) &=& y_0
\\
\vth(0) &=& \vth_0
\\
\vph_1(0) &=& 0
\\
\vph_2(0) &=& 0
\end{array}
\right. \quad \left\{
\begin{array}{lcl}
\dot{x}(0) &=& \dfrac{1}{2} R \cos \vth_0(\dot{\vph_1}_0 +
\dot{\vph_2}_0)
\\ \\
\dot{y}(0) &=& \dfrac{1}{2} R \sin \vth_0(\dot{\vph_1}_0 +
\dot{\vph_2}_0)
\\ \\
\dot{\vth}(0) &=& \dfrac{R}{L} \left(\dot{\vph_1}_0 -
\dot{\vph_2}_0\right)
\\ \\
\dot{\vph_1}(0) &=& \dot{\vph_1}_0
\\
\dot{\vph_2}(0) &=& \dot{\vph_2}_0
\end{array}
\right.
\end{eqnarray}
 with the only condition $\dot{\vph_1}_0 \ne
\dot{\vph_2}_0$, the motion of the system is given by
\begin{eqnarray}\label{moto_es1}
\left\{
\begin{array}{lcl}
{x}(t) &=& \dfrac12 L \dfrac{\dot{\vph_1}_0+
\dot{\vph_2}_0}{\dot{\vph_1}_0 -
\dot{\vph_2}_0}\left[\sin\left(\dfrac{R}{L} \left(\dot{\vph_1}_0 -
\dot{\vph_2}_0\right)t + \vth_0 \right) - \sin \vth_0 \right] +
x_0
\\ \\
{y}(t) &=& - \dfrac12 L \dfrac{\dot{\vph_1}_0+
\dot{\vph_2}_0}{\dot{\vph_1}_0 -
\dot{\vph_2}_0}\left[\cos\left(\dfrac{R}{L} \left(\dot{\vph_1}_0 -
\dot{\vph_2}_0\right)t + \vth_0 \right) - \cos \vth_0 \right] +
y_0
\\ \\
\vth (t) &=& \dfrac{R}{L} (\dot{\vph_1}_0 - \dot{\vph_2}_0)t
\\ \\
{\vph_1}(t) &=& \dot{\vph_1}_0 t
\\ \\
{\vph_2}(t) &=& \dot{\vph_2}_0 t
\end{array}
\right.
\end{eqnarray}
The linear and angular momenta equations for the system can be
written as
\begin{eqnarray}\label{sistema_generale_es1}
\left\{
\begin{array}{l}
(2m+M)\underline{{g}} + \underline{{\bf \Phi}}_{T_1} +
\underline{{\bf \Phi}}_{T_2} \= (2m+M) \underline{{\bf a}}_G
\\ \\
\overrightarrow{G{C_1}} \times m\underline{{g}} +
\overrightarrow{G{C_2}} \times m\underline{{g}} +
\overrightarrow{G{T_1}} \times \underline{{\bf \Phi}}_{T_1} +
\overrightarrow{G{T_2}} \times \underline{{\bf \Phi}}_{T_2}
\\ \\
\quad\quad\quad \= {\bf I}_{C_1}(\underline{\dot{\omega}}_1) +
\underline{{\omega}}_1 \times {\bf
I}_{C_1}({\underline{\omega}}_1) + m \overrightarrow{G{C_1}}
\times \underline{{\bf a}}_{C_1}
\\ \\
\quad\quad\quad\quad\quad + {\bf
I}_{G}(\dot{\underline{\omega}}_{rod}) +
{\underline{\omega}}_{rod} \times {\bf
I}_{G}({\underline{\omega}}_{rod})
\\ \\
\quad\quad\quad\quad\quad \quad\quad + {\bf
I}_{C_2}(\underline{\dot{\omega}}_2) + {\underline{\omega}}_2
\times {\bf I}_{C_2}({\underline{\omega}}_2) + m
\overrightarrow{G{C_2}} \times \underline{{\bf a}}_{C_2}
\end{array}
\right.
\end{eqnarray}
Taking into account the motion of the system (\ref{moto_es1}) and
introducing the orthonormal base $\{\underline{{\bf u}},
\underline{{\bf v}}, \underline{{\bf z}} \}$ with $\underline{{\bf
u}} = \dfrac{\overrightarrow{C_2C_1}}{L}, \underline{{\bf z}} =
\dfrac{\overrightarrow{T_1C_1}}{R}, \underline{{\bf v}} =
\underline{{\bf z}} \times \underline{\bf u}$, with obvious
notation we obtain
\begin{eqnarray}\label{reazioni_es1}
\left\{
\begin{array}{ccl}
\Phi_{1_z} &=& \dfrac12 \left[(2m+M)g - (3m+M)
\dfrac{R^3}{L^2}(\dot{\vph_2}^2_0 - \dot{\vph_1}^2_0) \right]
\\ \\
\Phi_{2_z} &=& \dfrac12 \left[(2m+M)g - (3m+M)
\dfrac{R^3}{L^2}(\dot{\vph_1}^2_0 - \dot{\vph_2}^2_0) \right]
\\ \\
\Phi_{1_v} &=& 0
\\ \\
\Phi_{2_v} &=& 0
\\ \\
\Phi_{1_u} + \Phi_{2_u} &=& - \dfrac12 (2m+M) \dfrac{R^2}{L}
(\dot{\vph_1}^2_0 - \dot{\vph_2}^2_0)
\end{array}
\right.
\end{eqnarray}
Note that, if the system leans on the horizontal plane, we must
add the requirement
\begin{eqnarray}\label{condizioni_esistenza_es1}
|\dot{\vph_1}^2_0 - \dot{\vph_2}^2_0| \le \dfrac{(2m+M)}{(3m+M)}
\dfrac{L^2}{R^2} \dfrac{g}{R}
\end{eqnarray}
since otherwise one between $\Phi_{1_z}$ and $\Phi_{2_z}$ becomes
negative (and this is not acceptable, because the system lifts
from the horizontal plane, and the initial assumptions of five
degrees of freedom is violated).

If (\ref{condizioni_esistenza_es1}) is fulfilled, then we can
chose for example $\Phi_{1_u} = \lambda$ and we find the reactions
$\underline{{\bf \Phi}}_{T_1}, \underline{{\bf \Phi}}_{T_2}$ as
functions of $\lambda$: Coulomb conditions (\ref{condizioni_CM})
then takes the form:
\begin{eqnarray}\label{condizioni_CM_es1}
\left\{
\begin{array}{l}
|\lambda| \le \mu_1 \, \dfrac12 \left[(2m+M)g - (3m+M)
\dfrac{R^3}{L^2}(\dot{\vph_2}^2_0 - \dot{\vph_1}^2_0) \right]
\\ \\
\left| \dfrac12 (2m+M) \dfrac{R^2}{L} (\dot{\vph_1}^2_0 -
\dot{\vph_2}^2_0) + \lambda \right|  \le \mu_2 \, \dfrac12
\left[(2m+M)g - (3m+M) \dfrac{R^3}{L^2}(\dot{\vph_1}^2_0 -
\dot{\vph_2}^2_0) \right]
\end{array}
\right.
\end{eqnarray}
In conclusion, the pure rolling of the disks can subsist if and
only  if (\ref{condizioni_esistenza_es1}) holds and there is a
$\lambda$ such that
\begin{eqnarray*}
\begin{array}{l}
\max \left\{- \dfrac12 \mu_1 \left[(2m+M)g - (3m+M)
\dfrac{R^3}{L^2}(\dot{\vph_2}^2_0 - \dot{\vph_1}^2_0) \right],
\right.
\\
\qquad\qquad\left. - \dfrac12 (2m+M) \dfrac{R^2}{L}
(\dot{\vph_1}^2_0 - \dot{\vph_2}^2_0) \right.
\\
\qquad\qquad\qquad\qquad\left. - \dfrac12 \mu_2  \left[(2m+M)g -
(3m+M) \dfrac{R^3}{L^2}(\dot{\vph_1}^2_0 - \dot{\vph_2}^2_0)
\right] \right\}
\\ \\
\qquad\qquad\qquad\qquad\qquad\qquad \le \lambda \le
\\ \\
\qquad\qquad \min \left\{ \dfrac12 \mu_1 \left[(2m+M)g - (3m+M)
\dfrac{R^3}{L^2}(\dot{\vph_2}^2_0 - \dot{\vph_1}^2_0) \right],
\right.
\\
\left. \qquad\qquad \qquad\qquad \dfrac12 \mu_2  \left[(2m+M)g -
(3m+M) \dfrac{R^3}{L^2}(\dot{\vph_1}^2_0 - \dot{\vph_2}^2_0)
\right] \right.
\\
\qquad\qquad \qquad\qquad\qquad\qquad \left. - \dfrac12 (2m+M)
\dfrac{R^2}{L} (\dot{\vph_1}^2_0 - \dot{\vph_2}^2_0)
 \right\}.
\end{array}
\end{eqnarray*}
\subsection{Example 3.}

\begin{figure}[h] \label{sfera}
\centering
\includegraphics[width=0.85\textwidth]{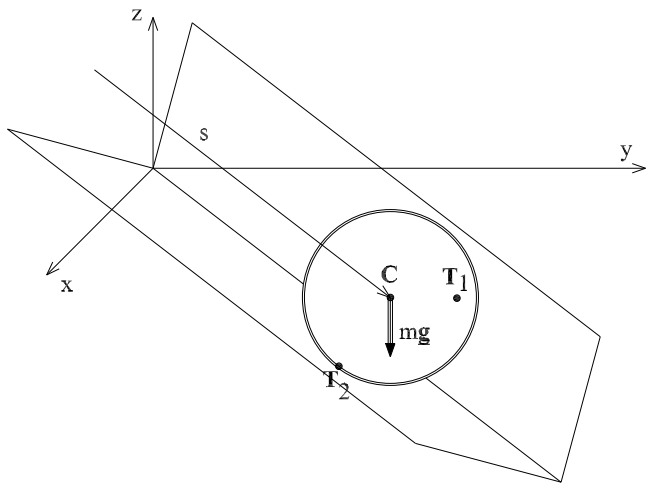}
\caption{Sphere on a V--groove}
\end{figure}

A mechanical system is formed by a sphere of mass $m$ and radius
$R$ leaned in an inclined V--groove whose walls are described by
the equations
\begin{eqnarray*}
\pi_1: 2x + y + z \=0 \, ; \quad \pi_2:  - x + y + z \=0
\end{eqnarray*}
We introduce an orthonormal base $\{ {\underline{\bf
k}}^{\perp}_{1}, {\underline{\bf k}}^{\perp}_{2}, {\underline{\bf
k}}^{\|} \}$ where ${\underline{\bf k}}^{\perp}_{1},
{\underline{\bf k}}^{\perp}_{2}$ are orthogonal to $\pi_1, \pi_2$
respectively and ${\underline{\bf k}}^{\|} = {\underline{\bf
k}}^{\perp}_{1}\times {\underline{\bf k}}^{\perp}_{2}$. The center
$C$ of the sphere is then determined by the vector $R
{\underline{\bf k}}^{\perp}_{1} + R {\underline{\bf
k}}^{\perp}_{2} - s {\underline{\bf k}}^{\|}$, where $s$ is the
distance of $C$ from a fixed plane orthogonal to ${\underline{\bf
k}}^{\|}$ (see Fig. \ref{sfera}). The rolling conditions in the
contact points $T_1, T_2$ determines the angular velocity of the
sphere in the form ${\underline{\omega}} = -
\dfrac{\dot{s}}{R}({\underline{\bf k}}^{\perp}_{1} -
{\underline{\bf k}}^{\perp}_{2})$ and the system has one degree of
freedom: the coordinate $s$.

The linear and angular momenta equations for the system can be
written as
\begin{eqnarray}\label{sistema_generale_es2}
\left\{
\begin{array}{l}
m\underline{{g}} + \underline{{\bf \Phi}}_{T_1} + \underline{{\bf
\Phi}}_{T_2} \= m \underline{{\bf a}}_C
\\ \\
\overrightarrow{T_1C} \times m\underline{{g}} +
\overrightarrow{T_1T_2} \times \underline{{\bf \Phi}}_{T_2} \=
{\bf I}_C(\underline{\dot{\omega}})  + m \overrightarrow{T_1C}
\times \underline{{\bf a}}_C
\end{array}
\right.
\end{eqnarray}
The projection of the angular momenta equation in the direction of
$\overrightarrow{T_1T_2}$ gives the equation of motion of the
sphere, that is $\ddot{s} = \dfrac{5\sqrt{2}}{18}g$. This relation
suffices to obtain from (\ref{sistema_generale_es2}) the
under--determined system of the reactions: if we decompose the
reactions along the basis introduced above
\begin{eqnarray*}
\underline{{\bf \Phi}}_{T_1} = \Phi_{1_N}{\underline{\bf
k}}^{\perp}_{1} + \Phi_{1_u}{\underline{\bf k}}^{\perp}_{2} +
\Phi_{1_v}{\underline{\bf k}}^{\|}
\\ \\
\underline{{\bf \Phi}}_{T_2}= \Phi_{2_u}{\underline{\bf
k}}^{\perp}_{1} + \Phi_{2_N}{\underline{\bf k}}^{\perp}_{2} +
\Phi_{2_v}{\underline{\bf k}}^{\|}
\end{eqnarray*}
the system takes the form
\begin{eqnarray}\label{reazioni_es2}
\left\{
\begin{array}{ccl}
\Phi_{1_v} &=& \dfrac{\sqrt{2}}{9} mg
\\ \\
\Phi_{2_v} &=& \dfrac{\sqrt{2}}{9} mg
\\ \\
\Phi_{1_N} + \Phi_{2_u} &=& \dfrac{1}{\sqrt{6}} mg
\\ \\
\Phi_{1_u} + \Phi_{2_N} &=& \dfrac{1}{\sqrt{3}} mg
\\ \\
\Phi_{2_N} + \Phi_{2_u} &=& \dfrac{1}{\sqrt{3}} mg
\end{array}
\right.
\end{eqnarray}
To analyze the parametric solution of the system we chose
$\Phi_{1_u} = \lambda mg$. In this case, and once again supposing
the sphere leaned on the groove, we must require the condition
$\lambda < \frac{1}{\sqrt{6}}$ since otherwise $\Phi_{1_N} < 0$
and the sphere comes off the groove. Conditions
(\ref{condizioni_CM}) take in this case the form
\begin{eqnarray}\label{condizioni_CM_es2}
\left\{
\begin{array}{l}
\lambda^2 + \dfrac{2}{81}  \le \mu_1^2 \,
\left(\dfrac{1}{\sqrt{6}} - \lambda \right)^2
\\ \\
\lambda^2 + \dfrac{2}{81} \le \mu_2^2 \, \left(\dfrac{1}{\sqrt{3}}
- \lambda \right)^2
\end{array}
\right.
\end{eqnarray}
with $\lambda < \dfrac{1}{\sqrt{6}}$. A straightforward minimum
computation for the functions on the left-hand side of
(\ref{condizioni_CM_es2}) shows then that the system can roll on
both the contact points if and only if
\begin{eqnarray}\label{condizioni_CM_es2_bis}
\left\{
\begin{array}{l}
 \mu_1 \ge \dfrac{2}{\sqrt{31}}
\\ \\
\mu_2 \ge \sqrt{\dfrac{2}{29}}
\end{array}
\right.
\end{eqnarray}

\section{Conclusions}
The procedure described in Sec. 2 in the case of two contact
points can be generalized to (multibody) systems with three or
more contact points (think for example of a ''steering tricycle''
formed by three vertical disks connected with three rods leaned on
the horizontal plane). Of course, an increase of the number of
contact points implies in general an increase of the technical
difficulties in practical applications. This is principally due to
the fact that Step 1 of the general procedure is not a
straightforward passage. The effective knowledge of the motion of
the system can be achieved only in some particular cases.
Insurmountable technical difficulties can arise both for
geometrical reasons (think of a convex rigid body moving in
contact with a surface, both having generic shapes with the only
requirement that the contact between rigid body and groove happens
in two points. For a more detailed discussion on the argument,
see, e.g. \cite{Hermans1995,BoriMama2002}), and/or for
computational reasons (even when the equations of motion of the
system are explicitly obtained, it could be hard to integrate them
to obtain the motion of the system). Nevertheless note that, as
pointed out by the examples in Sec. 3, not for all the systems the
explicit integration of the equations of motion is required.

A second remark is that the general procedure gives necessary
conditions such that the pure rolling subsists in all contact
points (conditions that become sufficient if we take into account
Step 4 of the procedure) but it does not give any information on
the behavior of the system if the pure rolling is not possible
even in a  single contact point. In fact, analogously to what
happens in the simple case of Ex. 1, in the instant when
(\ref{condizioni_CM}) stops to hold, the dynamics of the system
(for example, the number of degrees of freedom) changes abruptly.

To clarify this fact, suppose that, at the instant $t_1$ of the
study of the system of Ex. 2, a sudden variation of the friction
coefficient $\mu_2$ in the point $T_2$ (an oil spot on the plane?)
causes the invalidity of the second relation of
(\ref{condizioni_CM_es1}), while the first relation still holds.
Of course, even if we chose the assumption of Step 4 of the
procedure as a fixed point of our argument, we cannot suppose that
the system continues to roll in $T_1$ (and begins to slide in
$T_2$) since the beginning of sliding in $T_2$ can affect the pure
rolling behavior of the system in the point $T_1$. We must perform
a new analysis of the behavior of the system, possibly supposing
the system rolling in $T_1$ and sliding in $T_2$, we must
determine (if possible) the new equations of motion of the system
(with the additional difficulties of different friction laws in
the point $T_1$ and $T_2$ and possibly increased number of degrees
of freedom), the motion of the system, the new (parametric) system
of reactions acting on the system and then we can test the Coulomb
condition in the point $T_1$.

\end{document}